\documentclass[11pt,twocolumn]{article}
\usepackage{graphicx}

\begin{document}

\title{On the effects of a centralized computer routing and reservation system on 
   the electric vehicle public charging network. 
}
\author{ Thomas Conway 
\thanks{Contact: 
Dr. Thomas Conway
Lecturer, ECE Dept
University of Limerick
National Technology Park
Limerick, Ireland.
Tel +353 61 202628,
Email thomas.conway@ul.ie
}
\thanks{
This work has been submitted to the IEEE for possible publication. Copyright may be transferred without notice, 
after which this version may no longer be accessible.
}
}

\maketitle

\begin{abstract}
One solution to the limited range of battery electric vehicles is the 
provision of a public charging infrastructure to enable longer journeys.
This paper describes a simulation model of a centralized computer routing and reservation 
system based on the current charging infrastructure deployed (early 2016)
in Ireland using the Irish population density and a trip length
distribution. Monte Carlo simulations show quantitatively
the effects of EV on-board charger power rating and the advantages of a routing
and reservation systems on a country wide scale in terms of the number of electric
vehicles that can be supported. The effect of charge point fault
rates based on the currently deployed charging infrastructure is also assessed.
\end{abstract}

\section{Introduction}

   The advantages of electrified transportation are well known since the 1900's.
  Battery powered electric passenger cars and light commercial vehicles 
  are presently being manufactured and sold to the general public in many countries.  
  The technology of electric vehicles (EVs) is well developed and mature. 
  Modern battery electric vehicles can meet the needs of the majority of users most of the time.  
  However, the small percentage of trips that exceed the available range, present
  a stumbling block to their widespread adoption by consumers.  This
  {\it 'range anxiety'} \cite{ref:rangeanx1}\cite{ref:rangeanx2} needs to be 
  addressed if EV adoption rates are to increase.

   One possible solution is the deployment of a charging infrastructure,
  available to EV users, to allow recharging of the vehicle battery at 
  intermediate points during their trip \cite{ref:routeplanner1}. 
  Therefore, it is of considerable interest to evaluate the  performance of such infrastructure
  and determine its potential in addressing long trip requirement 
  of EV users.  
   Prior work using stochastic models network models \cite{ref:GaoStocasticModel}
  and recently intention aware routing models \cite{ref:ITSintentionaware}
  show improvements in journey times using prior history statistics.
  Deterministic central planning has been proposed previously with data
  presented for a grid road network with stations randomly
  deployed \cite{ref:centralcharging}.

   In this paper, the Republic of Ireland is taken as a case study, as
  already a comprehensive network of public charge points have been 
  deployed \cite{ref:EuropeEVchargepoints}.  A simulation model of the presently
  deployed charging infrastructure is developed in section~\ref{sec:model},
  based on the geographical population density, a trip length probability 
  distribution function and a routing algorithm that allows for reservation
  of charging points and minimization of travel time.  
  Monte Carlo simulations are run based on a specified number of EVs
  with metrics calculated to show the performance of the system on a
  countrywide scale.

  The results show the importance of the on-board charger power rating 
  which would be intuitively expected.  They also show the key importance
  of providing a charge point reservation systems in addition 
  to the physical charge points.  Such a reservation system together with 
  an optimizing routing algorithm is shown to provide a significant improvement
  in the number of EVs that can be supported under minimum average trip speed 
  specifications.

  While average trip speed is important, the concept of 'range anxiety' is
  really related to the chance of being stranded, i.e. running out of 
  battery energy and being unable to recharge.  In this paper, the effect
  of charge point faults is also considered.  This is the case of arriving
  at a charge point with a deeply depleted battery energy level only to
  find the charge point is not functional.   If it is not possible to
  travel to another charge point, then the EV is considered stranded and
  the user is unable to complete their trip.  The probability of this 
  occurrence must be comparable with current levels of trip failure, such
  as mechanical breakdown, if extensive adoption of EVs is to occur.

  In this paper, a system simulation model is described in 
  section~\ref{sec:model}. The results of  Monte Carlo simulations
  on this model are presented in section~\ref{sec:results}.

 %% FAULTS 

\section{ System Simulation Model \label{sec:model} }

The simulation model employed assumes that a specified number $N_{EV}$,
of EVs are deployed and that each one will make a trip, all starting at
the same time.   The start location of the trip is chosen from a 
geographical population density map of the country as described in 
section~\ref{sec:popdensity}.  The length of
the trip is randomly chosen from a trip length probability distribution 
function as developed in section~\ref{sec:trippdf}.
 The destination location is then chosen based on the population density map
of locations that are the chosen trip length distance from the 
start location.

Using a typical EV specification detailed in section~\ref{sec:EVspecs},
a routing algorithm is run for each trip.  If the trip length is less
than the available range, then no recharging is required and the
trip is assumed to be achievable with the normal vehicle speed.
Otherwise, the routing algorithm chooses a route using charge points
to ensure the trip can be completed.  The arrival time and charging
time at each charge point is calculated and a database entry made of
this information.  Subsequent trips being routed use this reservation database 
to ensure that any charge point is not allocated to more than one EV
at any given time.  As more trips are routed and charge points reserved,
it may become necessary for EVs to wait at a charge point, thus decreasing
the average trip speed for that vehicle.  The routing algorithm may
chose a longer distance trip through other charge points with less
waiting if the overall achievable trip time is less.
After processing $N_{EV}$ trips, the trip statistics are calculated.

\subsection{Population Distribution\label{sec:popdensity}}
\begin{figure}[htb]
  \centering
    \includegraphics[width = 0.5\textwidth]{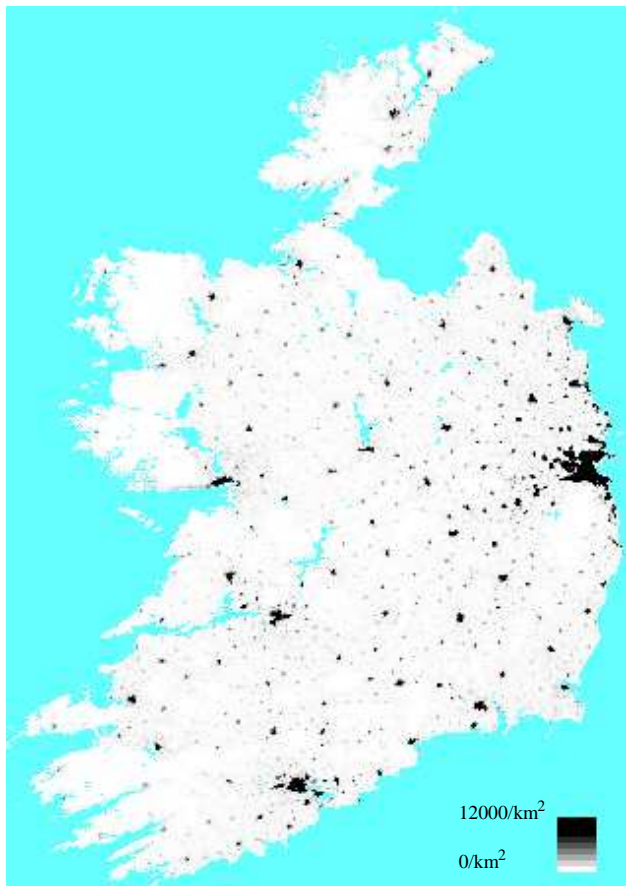}
  \caption{Population density map based on 2011 census data from \cite{ref:cop2011}}
  \label{fig:popdensitymap}
\end{figure}

A population distribution model  for the Republic of Ireland is developed  
based on data from the 2011 Irish census \cite{ref:cop2011}.  The data is used to
create a geographical map of the population density.
Fig.~\ref{fig:popdensitymap} shows the population density based on 
1 km by 1 km area blocks.

The 2011 Irish census further reports "1.36 million households having 
at least one  car".  This number is taken as the potential maximum
adoption of electric vehicle ownership for the purpose of the
developed model. Hence a 20\% electric vehicle adoption rate is
interpreted as $N_{EV} = 272000$ electric vehicles.  
The users of these vehicles are assumed, for the purpose of the model, to be distributed in the 
same manner as population density.  

\subsection{Journey Distribution\label{sec:trippdf}}
\begin{figure}[htb]
  \begin{center}
   \includegraphics[width = 0.5\textwidth]{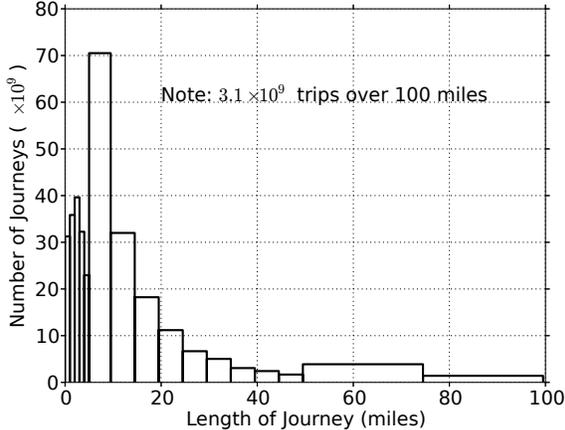}
  \end  {center}
  \caption{US journey distance distribution from \cite{ref:NHTS2009}.}
  \label{fig:NHTS2009raw}
\end{figure}

The distribution of journey distances is a key factor in the analysis
of electric vehicle usage models.  The Irish central statistics office
report that "On average, each private car travelled 16,736 kilometers 
in 2013" \cite{ref:cop2013}, but the distribution of journey distances is
not available.  However, an extensive survey by the US Federal Highway
Administration is available based on the 2009 National Household 
Travel Survey (NHTS).  Data extracted from this survey \cite{ref:NHTS2009},
provides the distribution shown in Fig.~\ref{fig:NHTS2009raw}. This data
shows an average journey length of 8.9 miles (14.2 km) per trip with less than
1\% of trips being over 100 miles (161 km).
The empirical probability density function in the journey length ($y$) in km 
\begin{equation}
%% 1.2059*kms*exp(-2.7733*(kms**0.33))
 p(y) = 1.2059 y  e^{-2.7733 y^{0.33}} 
\label{eqn:pdf}
\end{equation}
is developed based on this data and yields an average journey length of  16.7 km
with 1\% of trips over 161 km.
The annual travel distance of 16736 km indicates an average of  2.74 trips
per day.  

\subsection{Electric Vehicle Characteristic\label{sec:EVspecs}}

%%BATTERYCAPACITY = 24.0 # kWhr
%%AVGSPEED = 90.0 # kph
%%MAXRANGE = 120.0 # km
%%DCCHARGEMAX = 50.0 # kW
%%ACCHARGEMAX =  6.6 # kW
%%RANGEFACTOR = 0.85   # Account for non direct travel
%%MINRESERVE  = 0.20   # Plan always to be above this

While there are a number of electric vehicles available with different
characteristics, the parameters in table~\ref{tab:EVparams} are taken
as representative of a typical family sized electric car at the present time.
As a baseline scenario, it will be assumed that the user has the ability
to charge at their home and their work location to 100\% at the 6.6kW rate
using level 2 charging.

   \begin{table}[htb]
      \begin{center}
         \begin{tabular}{|c|c|}
            \hline
      Parameter &  Value \\
            \hline
            \hline
   Battery Capacity   &   24 kWhr \\
   Average Speed      &   90 kph \\
   Max Range          &   110 km \\
   DC charge rate (to 80\%)    &   45 kW  \\
   AC charge rate (to 100\%)        &  6.6 kW  \\
            \hline
         \end{tabular}
         \caption{Typical electric vehicle parameters.}
         \label{tab:EVparams}
      \end{center}
   \end{table}
With these characteristics, starting from a 100\% charge then
travelling until 20\% of the battery energy remains, a journey
distance of $ 110 km \times 0.8  = 88 km $ would be viable without
charging. Based on the distribution in Eqn.~\ref{eqn:pdf}, 
only about 2\% of journeys would require charging, en route.

For short trips, where no charging is required an average speed of 
90 kph, is assumed.  With a maximum charging rate of 45 kW, a 20\%  
to 80\% recharge time of 19.2 minutes is required and a distance of 66 km 
can be travelled between recharges.  At 90 kph, the time travelling between
charges is 44 minutes.  This results in a lower effective speed of 62.7 kph
if no waiting at charging facilities is assumed and the maximum charge power 
that the vehicle can take is available.

With 22 kW charging availability, the lower effective speed is 47.6 kph.

The effective speeds represent the limitation imposed by the charging requirement.
Speeds below these values represent limitations imposed by the finite 
charging infrastructure, a useful metric in assessing the 
quality of deployed infrastructure.

%% at 45KW  => 0.32 hr to charge 14.4KW   Vavg = 66/(0.7333333 + 0.32) = 62.7
%% at 22KW  => 0.6545454545454544 hr to charge 14.4KW  Vavg = 66/(0.7333+0.6545) = 47.6
%% or 70% and 53%

\subsection{Charge Point Allocation Algorithm\label{sec:chrgallocalgo}}

  The charge point allocation algorithm uses a database of available 
charge points consisting of their physical location, maximum power capability
and their reservation schedule.  A request for a journey route is handled
upon the arrival time of a reservation request.

 The allocation algorithm processes each reservation request by performing a 
breadth first search of all reachable charge points. The departure time from 
the charge point (including travel, charging and waiting times) is used as
as a metric.  To avoid infinite loops, any charge point considered is 
removed from the subsequent available list of charge point locations for that journey. 
Any consideration of a charge point that is within range of the final 
destination results in a viable route
for the trip.
The search paths are extended until all viable routes are found.
The best route in terms of the earliest arrival time is chosen.
Note that once any viable route is found, an overall arrival time is known.
Paths with a departure time later than the best time so far can be pruned
with no loss of optimality.  This achieves improved computational times by
avoiding extension of paths that can never be the optimum.

If a successful route is achieved, then the charge points on that route are 
reserved for the relevant times, otherwise a failure to route is declared.
 
The routing algorithm works on the basis of taking location to location (or point to point)
lengths ignoring the limitations of the road network.  A scaling
factor of $0.85$ is applied to all the vehicle ranges to provide for
some mitigation of the routing algorithm point to point assumption.
For example, with a fully charged battery and allowing the battery
energy to reach 20\%, the typical EV range from section~\ref{sec:EVspecs}
would be 88 km, but this is scaled to 88 km$\times0.85$ or 74.8 km 
as the maximum achievable point to point range before the first recharge
event. With a maximum allowable discharge of 20\% and recharging to 80\% 
at each recharge event, the maximum point to point
distance between charge points is  $ 110 \mbox{km} \times (0.8-0.2) \times 0.85$ or
56 km.

In a real deployment, a more realistic routing based on commercial navigation software 
could be employed \cite{ref:rangeanx1}, but this is beyond the scope of this study.  

\subsection{Fault Model\label{sec:faultsimalgo}}
While the reliability of the electric grid is generally very high in Ireland, there are many
reasons why public charge points may be non-operational at a particular time,
ranging from telematics issues, blocked access, vandalism, etc.  
In the worst case, the fault may be unknown to the charging utility or may just
have occurred when the EV driver arrives expecting to recharge their vehicle.
Using the charge point allocation algorithm of section~\ref{sec:chrgallocalgo}, the EV is
always expected to have a 20\% remaining capacity upon arrival at any charge point.

To evaluate the effect of charge point unavailability, simulations are run by
initially assuming all charge points are operational. The charge point allocation algorithm
is run with each trip needing recharging being allocated charge point which is reserved for
the corresponding EV.

A simple fault model is then assumed whereby a fraction of charge points are assumed to
be unavailable due to faults.  In this work, each individual charge point fault
is assumed to be independent.  The probability of a fault is denoted $p_f$.  

Any trip that includes a faulty charge point is stopped at the first
faulty charge point in its trip route.  The charge point allocation algorithm
is run with the start location being the first faulty charge point, the destination
location being  the original destination for that trip and the initial battery
capacity being the battery energy remaining on arrival at the first faulty charge point.
All faulty charge points are marked as non-operational during the algorithm run.
If it is not possible to reach any other operational charger, then the trip is considered
to have failed i.e. the EV is counted as stranded.

More detailed work on failure mechanisms is needed to evaluate the independent
fault assumption used here.  For example, circuit breaker events may disable
a bank of chargers deployed adjacent to each other.  However, the independent
fault assumption is used for simplicity in this study.

%%%  exhaustive minimum search? Maze Search
% maintenance, failure etc.
% priority, costing
% Margin 
% real-time updates, travel delays
% more accurate routing

%% diagram to illustrate

\section{Simulation Results\label{sec:results}}

\begin{figure}[htb]
  \begin{center}
    \includegraphics[width = 0.5\textwidth]{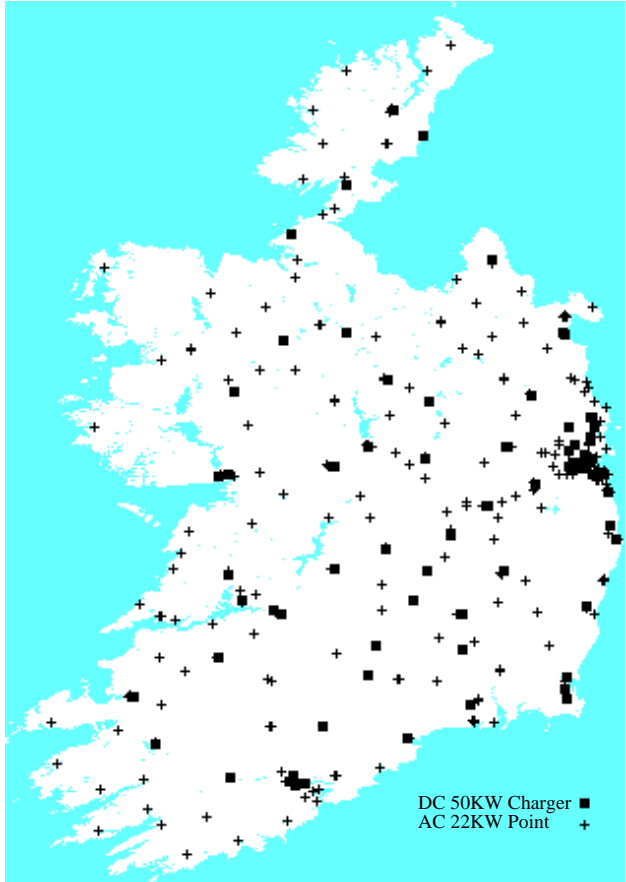}
  \end  {center}
  \caption{Location of public charge points on 11 Jan 2016}
  \label{fig:chargerJan11}
\end{figure}

The baseline scenario consists of taking the current distribution
of level 2 and level 3 (fast chargers) available in Ireland. 
It is assumed that it is possible to reserve their usage.
Based on data downloaded \cite{ref:esbkml} on 11 Jan 2016, there were 
72 DC chargers, 1 of which was not operational. There were 636 Type 2 AC 
charge points of which 49 were not operational.  Fig.~\ref{fig:chargerJan11} shows 
the location of these chargers. For simplicity, the DC chargers were
assumed to be 50 KW chargers units and the Type 2 AC charge points were 
assumed to be 22 KW 3 phase 230V units.  All the charge points are assumed
operational at the start of the simulation.

\subsection{Baseline Results}
\begin{figure}[htb]
  \begin{center}
    \includegraphics[width = 0.5\textwidth]{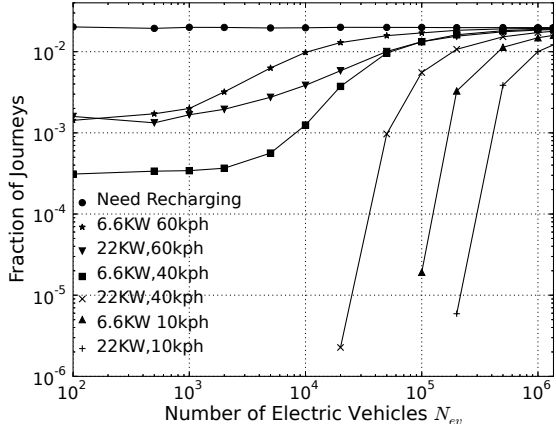} 
  \end  {center}
  \caption{Simulated baseline speeds for existing public charging infrastructure.}
  \label{fig:baselinesim}
\end{figure}

Two set of simulations are run.  The first assumes that the charging rate
at the 22 KW AC charge points is limited by the vehicle on board charger to 6.6 KW,
while the second assumes that the full 22 KW is available to charge the
vehicles battery.  The resulting data are shown in 
Fig.~\ref{fig:baselinesim}.  This figure shows the fraction of total trips
meeting various conditions.  In all cases, at least $10^7$ sample trips were
generated for each data point in the Monte Carlo simulations.

The first condition is that charging is required to complete the journey.
This happens in about 2\% of all the cases.  Such
a number would be expected based on the journey distribution as in 
section~\ref{sec:trippdf}.  

When charging is required the average speed is reduced due to the charging
time as well as waiting times.  The figure shows the fraction of total trips
that resulted in an average speed below 60 kph, 40 kph and 10 kph.
In these simulations, no trips were impossible.

From the figure, an average speed above 60 kph is not achieved in about
1 in 500 trips which are about 10\% of the 1 in 50 trips that require
recharging even with a very low number of vehicles.  As most of
the charge points are 22 KW, this result is not surprising.

Considering an average speed above 40 kph, all trips were able to exceed
this under the assumption that 22 KW on board charging was possible.
This was the case for supporting more than 10000 electric vehicles.
Naturally, the limitation of the 6.6 KW on board charging significantly
increases fraction of trips that fail to achieve 40 kph.
However, this data does show that the deployed infrastructure is
extensive; potentially supporting more than 10000 electric vehicles 
for trips over the whole country.  It also shows that employing 22 KW 
on board charging is a key factor in improving
the achievable average speed.
 
Above the 20000 electric vehicles, the limitation of the infrastructure
(waiting times) begins to dominate. Above 200000 electric vehicles,
many are beginning to hit average speeds below 10 kph. 

Choosing an acceptable probability of failing to achieve 40 kph as $10^{-4}$,
then the capacity of the currently deployed infrastructure would be
about 36000 vehicles.  This represents 2.6\% of the 1.36 million households 
having at least one car.

While an average trip speed of 40 kph seems low, it should be recalled that this
is a worst case value. For many non-professional drivers who take few
long distance trips, many of which may be leisure travel, a guarantee of 
this as worst case speed may be acceptable and enough to alleviate the
range anxiety associated with battery electric vehicles.

\subsection{Financial Costs}
The ability to support up to 36000 with the existing infrastructure
(assuming  22 KW on board charging with a routing and reservations system)
allows estimates of the financial cost per user to be calculated.  Based 
on the costs reported in \cite{ref:EuropeEVchargepoints}, the  average installation
costs of DC chargers and 22 KW AC charging posts were about 48K Euro and 12.5K Euro
respectively with annual maintenance costs of 6K Euro and 350 Euro.  Taking the
existing infrastructure of 72 DC chargers, 636 22 KW AC charging posts and
with an assumed lifespan of 20 years, then with 36000 users, the annual cost per user
would be 34 Euro per annum.  

If all the charge points were DC chargers, then the annual cost per user would
increase to 165 Euro per annum.

While these figures exclude overheads and the cost of the proposed 
routing and reservations system, they are reasonable in comparison to
the EV prices in the order of 30K Euro.

Presently the existing infrastructure has been subsidized on the basis of encouraging
EV adoption, but ultimately, the EV users would be expected to pay.  If financial charging
of EV users started when an EV adoption rate  of say 10\% of the potential 36000 users
was reached, the annual cost of 340 Euro per annum would be required.  
This amount would likely be acceptable to most users particularly if a guaranteed 
quality of service was provided.

%%TotalCost =  48000*72 + 12500*636  + 6000*20*72 + 350*20*636
%%UserAnnualAvgCost = TotalCost/(20*36000)

%%TotalCostALLDC =  48000*708+  6000*708*20
%%UserAnnualAvgCostALLDC = TotalCostALLDC/(20*36000)

\subsection{Effect of Reservations}
The prior simulations assumed that trips were reserved in advanced. The
allocated  charging times accounting for waiting times, to minimize the
overall journey time.  However, this is not currently available.
It is of interest to consider the impact
of such a reservation feature on average journey speeds.  Assuming 
22 KW on board charging the effect of such a  feature can be 
assessed.

\begin{figure}[htb]
  \begin{center}
    \includegraphics[width = 0.5\textwidth]{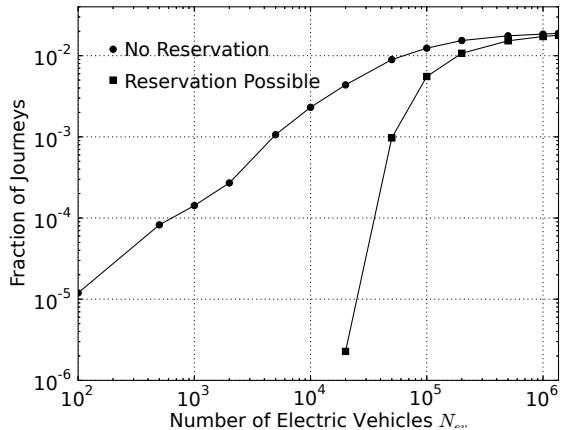}
  \end  {center}
  \caption{Simulated baseline speeds for existing public charging infrastructure with and without reservation provision for average trip speed above 40 kph.}
  \label{fig:booksim}
\end{figure}

 Fig.~\ref{fig:booksim} shows the data in the case of a reservation
algorithm that minimizes waiting times against the case where each
journey is planned based only on minimizing travel and charging time,
i.e. without consideration of waiting times due to other users.  
There is a severe deterioration in the fraction of trips that fail to
achieve an average speed above 40 kph.  This is the case even for relatively
small numbers of vehicles being electric.  It occurs because many users  
chose the same charge point, resulting in long waiting periods.  
Even with only 2000 vehicles, about 1\% of trips that need 
recharging fail to achieve the 40 kph level.  

With an acceptable probability of failing to achieve 40 kph as $10^{-4}$,
then the capacity of the currently deployed infrastructure with no
reservation system would be about 700 vehicles.  
Clearly there is an important need for a reservation infrastructure to
be deployed to maximize the utility of the physical charge point infrastructure.

The employment of a reservation and routing infrastructure also
allows for the implementation of a demand driven financial
costing model to allocate the financial cost of providing
the physical electrical charging infrastructure to EV users
\cite{ref:economicsofEV}.
  For example, fast DC chargers can be priced at a higher
rate than 22 KW AC charge points to reflect the additional costs
of the DC chargers. Indeed, some EV users may be happy 
to pay a higher rate for peak time use of fast DC charger while
others may be prepared to accept a longer trip time 
(e.g. using only 22 KW AC charge points) in return for lower costs.

\subsection{Fault Simulation}

In the case of charge point faults, the most serious problem is 
a vehicle being stranded and unable to complete its trip.
The probability of a vehicle being stranded in this manner is not 
related to the number of electric vehicles in the system, but only
the probability of a charge point fault $p_f$. 

With the installed base of 708 charge points, if 50 were non-operational (as
was the case on Jan 11, 2016),
then this would suggest a charge point fault probability of ${ 50\over 708}$ or
about 7\%.

Hence sequences of simulations are run for charge point fault
probabilities in the range of 1\% to 30\% as described in section~\ref{sec:faultsimalgo}.

The primary cause of stranded vehicles is arriving at a charge point
to find it non-operational and having insufficient battery energy left
to travel to another charger.  With the routing algorithm from 
section~\ref{sec:chrgallocalgo},  the worst case battery level on reaching
a charge point is set as 20\% capacity.  This corresponds to an available
point to point range of about 18 km.

Based on the charge point location distribution, there are 3 charge points
that have no neighboring charge points within this radius. Hence a 
fault at any of these would result in vehicles being stranded there
with a probability of order $p_f$.
Otherwise, at least two non-operational charge points would need to occur
as neighbors. This has probability of order $p_f^2$.
Thus, the probability of a stranded vehicle $p_{s}$ can be estimated
as
\begin{equation}
  p_{s}  \approx p_{c} p_f { 3\over 708}   +  O( p_f^2) + \ldots,
\label{eqn:pfaultx1}
\end{equation}
where $p_{c}$ is the probability of recharging being required ($\approx 2\%$)
and $\ldots$ represent third and  higher order terms in  $p_f$ .

As an example of improving the charge point infrastructure, three additional 
charge points were added to the model, one each co-located at the three identified
charge points with no neighbors in the 18 km  radius.  Simulations with these additional
charge points are also run.

It is also possible to modify the routing algorithm parameters to increase
robustness of the system.  The original reserve level for the battery energy 
was chosen as 20\% but increasing it to 28\% would ensure that sufficient 
reserve energy is available to avoid  being stranded in the case of a single
faulty charge point.

\begin{figure}[htb]
  \begin{center}
    \includegraphics[width = 0.5\textwidth]{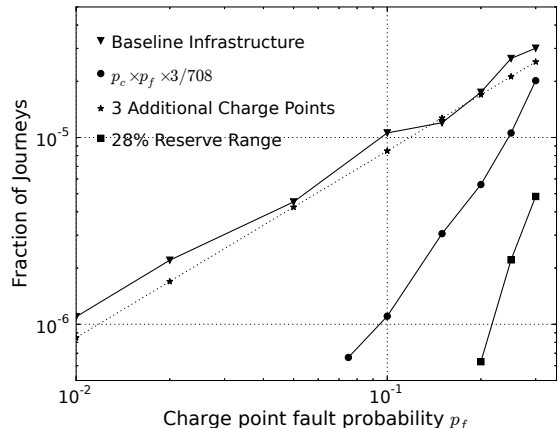}
  \end  {center}
  \caption{Fault simulation results.}
  \label{fig:faultsims}
\end{figure}

Fig.~\ref{fig:faultsims} shows the result of fault simulations.   
The baseline case of the 
existing infrastructure shows a stranding probability of about $10^{-5}$
for a fault probability of 10\%.  The baseline case is close to the
first term of Eqn.~\ref{eqn:pfaultx1}, indicating that the three charge points
identified are a significant source of stranded EVs in the model.

With the addition of just three additional charge points, Fig.~\ref{fig:faultsims} shows
almost a factor of ten improvement in the stranding probability to about $10^{-6}$
for a fault probability of 10\%.

Choosing an acceptable probability of being stranded of $10^{-6}$,
then the baseline case would require a fault probability of less than 1\%.
With the additional three charge points, the tolerable  fault probability
would be about 9\%.

The increase in the battery reserve energy level to 28\% with the baseline
infrastructure shows an even more significant improvement of the system robustness.
A fault probability of more than 20\% still achieves a probability of being stranded below
 $10^{-6}$.  However, increasing the reserve level reduces the maximum allowable distance 
between charge points.  The increase to 28\% resulted in a fraction of about $1.5\times10^{-5}$ 
trips not being possible to route in the first instance.   
In a real deployment a location dependent reserve level could be adopted which would address
this issue.

These results show that charge point fault probability and the 
charge point location distribution are key factors
in the stranding probability of EVs for long trips using a recharge
infrastructure.  
The robustness of the recharge infrastructure
can be increased by adding redundancy at existing charge points, even if they
are low power charge points just to reduce the probability of stranded vehicles.
The use of a routing and reservation system can also
significantly improve the system resilience to charge point faults.  For example
the allocation of a higher battery energy reserve when high risk charge points
are being used can significantly improve the system robustness.

\section{Conclusions}

In this paper, a model of the complete charge point infrastructure deployed (early 2016)
in Ireland is built using the Irish population density. The assumed trip length
probability density function is based on a US survey (as this data was not available 
for the Irish case).  The population density and trip length distribution are used
to create a set of trips based on the number of EVs assumed present. These trips
are then routed through the deployed charge points when the trip length exceeds
the range of a typically EV presently available.

The results of these simulations show that with the typically EV, that the deployed
charge point infrastructure is extensive and can support electrified travel across
the whole country.  With the majority of charge points being 22kW AC sources, the
effect of the on-board charger power rating is a limiting factor in the vehicle.
Manufactures are working on this \cite{ref:integcharge}. At least a 22kW 
power rating appears desirable.

  The second key factor is the provision of a routing and reservation system, which
is not presently available to EV users in Ireland.  Without this, the number
of EVs that the system can support is limited.  As measured by average trip speed, 
even with the assumption of 22kW on-board charger power ratings, the present
infrastructure could potentially support about 36000 EVs based on achieving an average trip speed
below 40 kph with a probability of $10^{-4}$. This is under the assumption of
a routing and reservation system as described in section~\ref{sec:chrgallocalgo}.
Without such a system, the equivalent number support is less than 1000.

The third factor that needs to be accounted for is the effect of charge point
faults.  The worst case scenario for the EV user is the chance of being
stranded at a faulty charge point, thus being unable to complete the journey
at any speed.  Using a fault model that assumes faults in each charge point
are independent, the system simulation can be used to assess the impact of
faults rates on the fraction of EVs being stranded.  The simulation results
show the importance of charge point distribution with low fault rates 
to reduce the probability of stranding EV and the ability of an intelligent
routing algorithm to improve the robustness of the system to charge point
faults.

Overall, the simulation model and results in this paper show quantitatively
the effects of EV on board charger power rating, the major advantage of a routing
and reservation on a country wide scale, and the effect of charge point fault
rates based on a currently deployed charging infrastructure.

 %Provides solution to the range anxiety
 %Scalable
 % Chamelion charger!
 %Cost borne by benificaries
 %Major Up from Government investment not require


\begin{thebibliography}{1}

\bibitem{ref:rangeanx1}
M. Neaimeh, G. A. Hill, Y. Hubner and P. T. Blythe. Routing systems
to extend the driving range of electric vehicles. Intelligent Transport
Systems, 7(3), pp. 327-336, 2013

\bibitem{ref:rangeanx2}
U.S. National Energy Technology Laboratory. Assessment of Future
Vehicle Transportation Options and Their Impact on the Electric Grid.
Report DOE/NETL-2010/1466, 2011

\bibitem{ref:routeplanner1}
L. Bedogni, L. Bononi, A. D'Elia, M. Di Felice, M. Di Nicola and T. S. Cinotti, 
Driving without anxiety: A route planner service with range prediction for the electric vehicles, 2014 International Conference on Connected Vehicles and Expo (ICCVE), Vienna, 2014, pp. 199-206.

\bibitem{ref:GaoStocasticModel}
S. Gao, 
Modeling Strategic Route Choice and Real-Time Information Impacts in Stochastic and Time-Dependent Networks, 
IEEE Transactions on Intelligent Transportation Systems, 
vol. 13, no.3 pp. 1298-1311, Sept 2012

\bibitem{ref:ITSintentionaware}
M. M. de Weerdt, S. Stein, E. H. Gerding, V. Robu and N. R. Jennings, 
Intention-Aware Routing of Electric Vehicles, 
IEEE Transactions on Intelligent Transportation Systems, 
vol. 17, no.5 pp. 1472-1482, May 2016


\bibitem{ref:centralcharging}
H. Qin and W. Zhang, Charging scheduling with minimal waiting in a
network of electric vehicles and charging stations, in Proc. 8th ACM Int.
Workshop Veh. Inter-Netw., 2011, pp. 51-60.


%doi: 10.1109/ICCVE.2014.7297541
\bibitem{ref:EuropeEVchargepoints}
P. R. LeGoy and G. J. M. Buckley, Low voltage grid connections for Electric Vehicle Infrastructure in Europe, 2012 IEEE Power and Energy Society General Meeting, San Diego, CA, 2012, pp. 1-8.



\bibitem{ref:cop2011}
  Central Statistics Office, 
COP2011\_Grid\_ITM\_IE\_1Km.zip URL:\\
{\tt\small http://www.cso.ie/en/census/ census2011griddataset}
accessed Dec 17, 2015

\bibitem{ref:cop2013}
  Central Statistics Office,  URL:\\
{\tt\small
http://cso.ie/en/releasesandpublications/ep/
p-tranom/transportomnibus2013/roadtraffic/
roadtrafficvolumes/}
accessed Dec 19, 2015


\bibitem{ref:NHTS2009}
U.S. Department of Transportation, Federal Highway Administration, 2009 National
 Household Travel Survey. URL: http://nhts.ornl.gov.
accessed Dec 19, 2015

\bibitem{ref:esbkml}
URL: https://www.esb.ie/electric-cars/kml/all.kml
accessed Jan 11, 2016

\bibitem{ref:economicsofEV}
E. Gerding, S. Stein, V. Robu, D. Zhao, and N. R. Jennings, Two-sided
online markets for electric vehicle charging, in Proc. Int. Conf. Auton.
Agents Multiagent Syst., 2013, pp. 989-996.



\bibitem{ref:integcharge}
I. Subotic, N. Bodo and E. Levi, An EV Drive-Train With Integrated Fast Charging Capability, IEEE Transactions on Power Electronics, vol. 31, no. 2, pp. 1461-1471, Feb. 2016.
\end{thebibliography}
\end{document}